\documentclass[trackchanges,twocolumn]{aastex631}

\newcommand{\V}[1]{\mathbf{#1}} 

\newcommand{\uofa}{\affiliation{University of Arizona, Tucson, AZ, USA}}
\newcommand{\swri}{\affiliation{Space Science and Engineering, Southwest Research Institute, San Antonio, TX, USA}}
\newcommand{\lesia}{\affiliation{LESIA, Observatoire de Paris, Meudon, France}}
\newcommand{\chch}{\affiliation{Charles University, Faculty of Mathematics and Physics, V Hole\v{s}ovi\v{c}k\'{a}ch 2, 180 00 Prague 8, Czech Republic}}

\received{August 10, 2021}
\revised{October 4. 2021}
\accepted{October 14. 2021}

\shorttitle{\title{Ion-Driven Instabilities as Observed by Helios}}
\shortauthors{Martinovi\'c et al.}

\begin{document}

\title{Ion-Driven Instabilities in the Inner Heliosphere I: Statistical Trends }

\correspondingauthor{M. M. Martinovi\'c}
\email{mmartinovic@arizona.edu}

\author[0000-0002-7365-0472]{Mihailo M. Martinovi\'c}
\uofa
\lesia

\author[0000-0001-6038-1923]{Kristopher G. Klein}
\uofa

\author[0000-0003-4247-4864]{Tereza {\v{D}}urovcov{\'a}}
\chch

\author[0000-0001-6673-3432]{Benjamin L. Alterman}
\swri


\begin{abstract}

Instabilities described by linear theory characterize an important form of wave-particle interaction in the solar wind. 
We diagnose unstable behavior of solar wind plasma between 0.3 and 1 au via the Nyquist criterion, applying it to fits of $\sim1.5$M proton and $\alpha$ particle Velocity Distribution Functions (VDFs) observed by \emph{Helios I} and \emph{II}. 
The variation of the fraction of unstable intervals with radial distance from the Sun is linear, signaling a gradual decline in the activity of unstable modes. 
When calculated as functions of the solar wind velocity and Coulomb number, we obtain more extreme, exponential trends in the regions where collisions appear to have a notable influence on the VDF. 
Instability growth rates demonstrate similar behavior, and significantly decrease with Coulomb number. 
We find that, for a non-negligible fraction of observations, the proton beam or secondary component might not be detected due to instrument resolution limitations, and demonstrate that the impact of this issue does not affect the main conclusions of this work. 
\end{abstract}



\section{Introduction} \label{sec:intro}

Instabilities, driven by departures from local thermodynamic equilibrium (LTE), are frequently credited with affecting the behavior of rapidly evolving plasma systems, e.g.  the expanding solar wind \citep{Matthaeus_2012_JTurb}.
To quantify these departures, the underlying charged particle Velocity
Distribution Functions (VDFs) are typically modelled as bi-Maxwellians, having anisotropic temperatures $T_{\perp,j}$ and $T_{\parallel,j}$ with respect to the local magnetic field $\V{B}$, relative field-aligned drifts between each pair of constituent VDF components $i$ and $j$ being $\Delta v_{i,j}=(\V{V}_i-\V{V}_j)\cdot \V{B}/|\V{B}|$, and temperature disequilibrium between species $T_i \neq T_j$ \citep{Marsch_1982}. 
These anisotropies, drifts, and disequilibrium serve as distinct sources of free energy capable of driving the growth of a number of distinct unstable solutions (see e.g, Section 5 of \cite{Verscharen_2019_LRSP}). 
The presence of multiple free-energy sources makes it difficult to determine which subset of sources drives a given instability; parametric models accounting for a single source of free energy---e.g. the temperature anisotropy of a single population \citep{Gary_1997_JGR,Yoon_2017_RvMPP}--- do not account for the diminishment or enhancement of predicted linear growth rates associated with the introduction of other departures from LTE--- e.g. relative drifts of proton beams \citep{Daughton_1998_JGR,Woodham_2019_ApJ,Liu_2021_arXiv}, helium \citep{Podesta_2011_ApJ_alpha,Verscharen_2013_ApJ_alpha,Bourouaine_2013_ApJ}, or their combined effects \citep{Chen_2016_ApJ}.
To account for these effects, previous studies implemented Nyquist's instability criterion \citep{Nyquist_1932,Klein_2017_JGRA} on limited sets of in situ measurements from the \emph{Wind} \citep{Klein_2018}, \emph{Parker Solar Probe} \citep{Verniero_2020_ApJS,Klein_2021_ApJ}, and \emph{Helios} \citep{Klein_2019_ApJ} missions, finding that a majority of intervals were unstable, that the kinds of waves driven unstable were very sensitive to the model used to describe the VDF, and that inclusion of multiple ion populations could both enhance or diminish the predicted growth rates.

In this work, we apply the Nyquist criterion to a recent reprocessing of the VDF measurements that provides fits for a proton core, proton beam, and helium ($\alpha$) component for nearly the entirety of both \emph{Helios I} and \emph{II} missions \citep{Durovcova_2019_SoPh}. 
Details of the method are given in Section \ref{sec:data}.
Due to \emph{Helios} I1a and I1b instruments \citep{Schwenn_1975} limitations in resolution, range and field of view \citep{Helios_BlueBook_1981,Marsch_1982}, the beam population is not always detected. 
In general, mischaracterization of the beam as a part of the core can lead to significant variations in prediction of modes excited by the VDF \citep{Klein_2021_ApJ}. 
For example, a stable VDF consisted of an almost isotropic core and mildly shifted isotropic beam could potentially be fitted as a single population with $T_{\parallel,c} \gg T_{\perp,c}$, highly susceptible to firehose (FH) instability. 
To ensure this issue does not have a misleading effect on our analysis, we detail the procedure of diagnosing insufficiently well resolved observations in Section \ref{ssec:Data_Caveats}, and remove them from consideration of the results. 

We find that the traditional way of organizing \emph{Helios} observations over radial distance \citep{Matteini_2007_GRL,Hellinger_2011,Hellinger_2013} does not provide a complete description of the evolution of linear instabilities, neither in terms of their occurrence rate nor the growth rate. 
The obtained linear trend for both parameters turns out to be a result of a mixture of varying solar wind parameters at each distance. 
On the contrary, organization of the results with respect to the solar wind bulk velocity $v_{sw}$ and, even more significantly, the Coulomb number $N_C$, reveals a specific region in which either speed or collisionallity seem to be main candidates for the solar wind parameter that governs the behavior of instabilities. 
The Coulomb number is a dimensionless measure that characterizes the VDF relaxation due to Coulomb collisions. 
It is defined as $N_{C(cc)} = \nu_{cc} r / v_{sw,c}$, where $\nu_{cc}$ is the Coulomb self-collision frequency among core protons (see  \cite{Hernandez_1987,Kasper_2017} for details). 
We do note that these two quantities are connected due to $\nu_{cc}$ scaling with the proton temperature, which is in turn correlated with $v_{sw}$ \citep{Elliott_2012_JGRA,Perrone_2019_MNRAS,Maksimovic_2020_ApJS}. 
We discuss the validity of these statistics and their implications to the general description of the instability behavior in the solar wind in Sections \ref{ssec:Radial_Trends} and \ref{sec:discussion}. 


\section{Data and Methodology}
\label{sec:data}


\subsection{Models for Ion Distributions from Helios I and II}
\label{ssec:data}

The previous application of the Nyquist criterion to \emph{Helios} observations focused entirely on selected fast-wind streams with good fits to the proton \citep{Stansby_2018_SolPh} and helium \citep{Stansby_2019_A&A} components, approximately $45,000$ intervals in total \citep{Klein_2019_ApJ}, while entirely neglecting proton beams.

In this work, we apply the same analysis method to $1,480,214$ intervals, from the \emph{Helios I} and \emph{II} missions, which also include the reprocessed version of intervals used by \cite{Klein_2019_ApJ}. 
From this database, $49,316$ (3.3\%) entries are excluded to remove observations of Coronal Mass Ejections \citep{Wang_2005_JGRA}. 
Further on, we remove $3,500$ (0.2\%) cases from further consideration due to the dispersion solver (Section \ref{ssec:analysis}) encountering numerical precision issues, and also $14,620$ (0.98\%) cases when the predicted growth rates could not be accurately described with linear physics (see Section \ref{ssec:analysis}). 
Fits assume bi-Maxwellian VDFs for the proton core, proton beam, and $\alpha$ component, and are described in \cite{Durovcova_2019_SoPh}. 
By not restricting our study to fast solar wind, we aim to remove selection bias---focusing mostly on faster, and therefore hotter and less collisionally processed solar wind, is naturally biased towards more unstable plasma, and has produced discrepancy in fractions of unstable intervals found by \cite{Klein_2019_ApJ} for \emph{Helios} and \cite{Klein_2018} for \emph{Wind} at 1 au. 
In this paper, we aim to provide a more comprehensive picture of the role instabilities play in all kinds of solar wind as it expands in the inner heliosphere.

The number of intervals with identified proton core, proton beam, and
$\alpha$ components are listed in Table \ref{tab:params}. 
Proton beams and $\alpha$ populations were distinguished in $843,224$ (59.7\%) and $744,609$ (52.7\%) of the fits. 
It is important to note that both populations could be completely or partially undetected in some of the fits\footnote{The I1a instrument could take 144 records over its entire angular domain (9 elevation and 16 azimuth channels) and its energy-to-charge range was covered by 32 channels. However, this exceeded the telemetry limits and data reduction had to be applied. In the normal data mode, the channel with the maximum count rate was found. In the next measurement cycle, only a limited number of channels around this maximum are recorded (9 energy, 5 azimuth and 5 elevation channels). These 9 energy channels are designed to also cover the helium core part of the VDF, making the resolution sparse. This complicates the ability to identify the proton beam. The $\alpha$ beam, which occurs at higher energy-to-charge ratios and has very low abundance, is almost never detected.}, but due to different reasons. 
The beam population can be misinterpreted as part of the core if it has low density\footnote{Following notation in \cite{Durovcova_2019_SoPh}, the VDF components are labeled in such a way that, by definition, $n_b / n_c \leq 1$, while the drift velocity of the beam can be negative.} $n_b / n_c \ll 1$ and/or the relative drift velocity between the two populations is small $\Delta v_{b,c} / v_{Ac} \ll 1$. 
Here, the core-proton Alfv\'en velocity defined as $v_{Ac} = B/\sqrt{\mu_0 n_c m_p}$, where $\mu_0$ is magnetic permeability of vacuum, and $n_c$ and $m_p$ are proton core density and mass, respectively. 
This type of under-detection is central technical issue of our data analysis and is discussed in detail in Section \ref{ssec:Data_Caveats}. 
On the other hand, the helium population is shifted in the instrument frame due to doubled mass-to-charge ratio compared to protons. 
This effect can cause the helium beams, and rarely even parts of the $\alpha$ bulk population to be out of the instrument field of view, or to be impossible to resolve due to lower instrument resolution at high energies. 
Therefore, the fits performed by \cite{Durovcova_2019_SoPh}, as well as in the fits by \cite{Stansby_2019_A&A} used for linear stability analysis by \cite{Klein_2019_ApJ}, represent $\alpha$ particles with a single shifted bi-Maxwellian. 
A direct consequence of this approach is that majority of the measurements yield $T_{\perp,\alpha} / T_{\parallel,\alpha} \leq 1$. 
Similar behavior is reported for protons if the core and beam populations were not separated \citep{Marsch_1981_JGR,Marsch_1982}, where core and moment proton anisotropy are significantly different when beams are present. 
While recent analysis of \emph{Parker Solar Probe (PSP)} observations \citep{Klein_2021_ApJ} demonstrated that treating protons as a single population could fail to correctly identify the beam-induced wave modes, the presence of which can be confirmed by electromagnetic field measurements \citep{Vech_2021_AA}, there is no such study dealing with $\alpha$ particles. 
New observations of the secondary helium population by \emph{PSP} \citep{McManus_2020_AGU} have the potential to provide precise measurements necessary for detailed analysis of the effect of these structures on plasma linear stability.

\begin{deluxetable*}{|c|c||cccc|}
\label{tab:params}
\tablecaption{Number of intervals with characterized proton core, proton beam, and $\alpha$ components and their percentage of the total data set, as well as the number of intervals classified as supporting an unstable mode with growth rates greater than $10^{-4}$, $10^{-3}$, and $10^{-2} \Omega_p^{-1}$, with cumulative percentages.}
\tablehead{
\colhead{} & \colhead{Total} & \colhead{core} & \colhead{core + beam} & \colhead{core + $\alpha$} &
\colhead{core + beam + $\alpha$} \\
\colhead{\#} & \colhead{1,412,778} & \colhead{315,755 (22.3\%)} & \colhead{352,414 (24.9\%)} & \colhead{253,779 (18.0\%)} & \colhead{490,830 (34.7\%)}
}
\startdata
    \hline
    $\gamma^{\textrm{max}}/\Omega_p > 10^{-4} $  & 630,540 (44.6\%) & 60,513 (19.2\%) & 215,223 (61.1\%) & 74,452 (29.3\%) & 280,352 (57.1\%)\\
    $\gamma^{\textrm{max}}/\Omega_p > 10^{-3} $  & 510,916 (36.2\%) & 47,115 (14.9\%) & 182,875 (51.9\%) & 47,911 (18.9\%) & 233,015 (47.5\%)\\
    $\gamma^{\textrm{max}}/\Omega_p > 10^{-2} $  & 316,825 (22.4\%) & 26,655 (8.4\%) & 122,617 (34.8\%) & 16,350 (6.4\%) & 151,203 (30.8\%)\\
    \hline
\enddata
\end{deluxetable*}

\subsection{Instability Analysis}
\label{ssec:analysis}

The Nyquist instability criterion \citep{Nyquist_1932}, using the hot
plasma dispersion relation for an arbitrary number of relatively drifting bi-Maxwellian components as calculated by the Plasma in a Linear Uniform Magnetized Environment (\texttt{PLUME}) numerical dispersion solver \citep{Klein_2015_PhPl}, has been described in detail in previous publications \citep{Klein_2017_JGRA}. 
This method determines the stability of a linear system of equations through a conformal mapping of the contour integral of the dispersion relation $\mathcal{D}(\omega,\V{k},\mathcal{P})$ over the upper-half of the complex frequency plane, with the real ($\omega_{\textrm{r}}$) and imaginary ($\gamma$) components of the frequency $\omega = \omega_\textrm{r} + i\gamma$ representing the oscillation and growth or damping rates respectively. 
This integral counts the number of normal mode solutions that are unstable, having $\gamma>0$, for a specific wavevector $\V{k}$ and set of dimensionless parameters $\mathcal{P}$. 
Iterating this process for multiple contours with increasing values of $\gamma$ enables the determination of the maximum growth rate and associated characteristics of the fastest growing mode supported by a particular $\V{k}$. 
We have set $\gamma = 10^{-4} \Omega_p$ as the minimum growth rate for a wavevector to be considered unstable, where proton gyrofrequency is given as $\Omega_p = e_c B / m_p$, $e_c$ being the elementary charge. 
We choose this particular minimum value of $\gamma$, as for lower values the growth time of the unstable mode becomes long enough for other physical effects---such as propagation, reconnection, or turbulence---to start notably changing the VDF parameters \citep{Livi_1987_JGR,Kasper_2017}.
We repeat this process over a log-spaced grid in wavevector space $k_\perp\rho_c \in [10^{-3},3]$ and $k_\parallel\rho_c \in [10^{-2},3]$, enabling the determination of the fastest growing mode for ion-scale wavevectors given a particular observed set of parameters $\mathcal{P}$.
This set is composed of two global reference parameters
\begin{equation}
  \mathcal{P}_{0}=\left(\beta_{\parallel,c},\frac{w_{\parallel,c}}{c}
  \right)
\end{equation}
where we define the thermal-to-magnetic pressure ratio $\beta_{\parallel,j}= 2 \mu_0 n_j k_b T_{\parallel,j}/B^2$, the speed of light $c$, the Boltzmann constant $k_b$, the thermal velocity $w_{\parallel,j}=\sqrt{2 k_b T_{\perp,j}/m_j}$, and gyroradius $\rho_j = m_p w_{\parallel,j} / e_c B$.
We also consider six parameters for each component $j$
\begin{eqnarray}
  \mathcal{P}_{j}=
  \left(\frac{n_j}{n_c},\frac{T_{\perp,j}}{T_{\parallel,j}},
  \frac{T_{\parallel,j}}{T_{\parallel,c}}, \frac{\Delta
    v_{j,c}}{v_{Ac}}, \frac{m_j}{m_p}, \frac{q_j}{q_p} \right).
\end{eqnarray}

For this study, we treat the electrons as a single isotropic distribution with density and drift velocity necessary to enforce quasi-neutrality and zero net current. 
The number of elements in $\mathcal{P}$ thus varies depending on the number of characterized ion components.

For a given $\mathcal{P}$, we calculated $\gamma/\Omega_p$ using the Nyquist method as well as $\gamma^{\textrm{max}}/\Omega_p$ over the entire wavevector range and the associated $\omega_{\textrm{r}}^{\textrm{max}}/\Omega_p$, $\mathbf{k}^{\textrm{max}} \rho_c$, $\theta_{kB}^{\textrm{max}}$, and electromagnetic eigenfunctions of the most unstable mode. 
The number of intervals with growth rates in excess of $10^{-4}, 10^{-3},$ and $10^{-2} \Omega_p^{-1}$ are listed in Table~\ref{tab:params}, both for the entire database and segregated as a function of the identified ion components. 
Slightly less than half of the intervals in the total database support at least a weakly growing mode, with nearly a quarter supporting a relatively rapidly growing mode, with $\gamma^{\textrm{max}} >10^{-2} \Omega_p^{-1}$. 
There is significant variation between the intervals with and without proton beams, as well as if an $\alpha$ component has been identified. 
Nearly a third of intervals with an identified proton beam component have $\gamma^{\textrm{max}} >10^{-2} \Omega_p^{-1}$, while less than 10 \% of intervals without proton beams have such strongly growing modes.
As mentioned in Section \ref{ssec:data}, for $\gamma^{\textrm{max}}/\omega_\textrm{r}^{\textrm{max}} > 0.3$ (0.98\% of all measurements) we do not consider the wave to be a small linear perturbation and exclude these intervals.


\section{Inferred Stability As A Function of Speed, Distance, and Coulomb Number}
\label{sec:results}

\begin{figure*}
  \includegraphics[width=\textwidth]{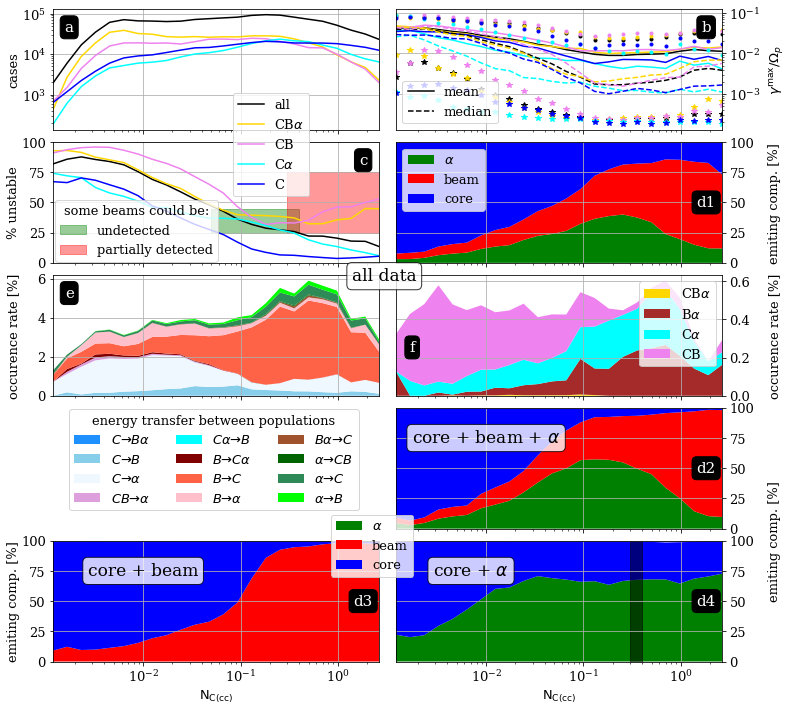}
  \caption{Overview of unstable mode behavior and VDF components that emit the most power as a function of $N_{C(cc)}$. 
  The means and medians of $\gamma^{\textrm{max}}/\Omega_p$ on \emph{panel (b)} are calculated considering only unstable cases, while 10\% and 90\% percentiles are given by stars and dots, respectively. 
  \emph{Panels (d)} are showing percentages of the component that emits the largest amount of power for unstable cases. 
  \emph{Panel (e)} illustrates the occurrence of specific intervals where one component emits power, at least 50\% of which is then absorbed by another population. 
  On the contrary, the \emph{panel (f)} shows the cases where two, or rarely even all three ion components have power emitting within 50\% of each other. 
  Red, green, and black shaded areas are used for beam detection tests described in Section \ref{ssec:Data_Caveats}.}
  \label{fig:overview_N_C}
\end{figure*}

The summary of our findings is given on Figure \ref{fig:overview_N_C}. 
The results are presented in term of the Coulomb number, which we pick as an initial referent parameter as it measures traits of both solar wind speed and radial distance, and only through their combined effects we were able to determine the parts of phase space that potentially have instrumental effects affecting VDF fits. 
Therefore, before we address the behavior of instabilities with respect to plasma parameters, it is first necessary to understand the consequences of instrument limitations regarding the characterization of the core, beam and helium populations and, more importantly, their effects on our linear stability analysis. 

\subsection{Observational issues of I1 instrument}
\label{ssec:Data_Caveats}

The trends on the \emph{panels (b)} and \emph{(c)} of Figure \ref{fig:overview_N_C} are matching with intuitive expectations - collisionally young solar wind is strongly unstable, before both occurrence rate and intensity of instabilities start declining. 
However, unexpected behavior at moderate and large $N_{C(cc)}$ can be noticed for two regions in the phase space. 
First, in rare occasions when the beam is detected in the VDF at large $N_{C(cc)}$, the interval seems to commonly be very unstable, producing significant increase in fraction of unstable intervals and maximum growth rate (red rectangle). 
Second, intervals at moderate $N_{C(cc)}$ where only core and $\alpha$ populations were detected (green rectangle), show a notable increase in the number of unstable VDFs and  $\gamma^{\textrm{max}}/\Omega_p$. 

The first issue, related to beams at large $N_{C(cc)}$, is rooted in the difficulties of beam detection. Three properties characterize the significance of proton beams; $n_b / n_c$, $\Delta v_{b,c} / v_{A,c}$ and $T_{\perp,b} / T_{\parallel,b}$. A comparison of these quantities at large $N_{C(cc)}$ and r demonstrated that a beam was resolved only if at least one of the three was large. 
Otherwise, the beam population was indistinguishable from the core. 
Since a large value of any of the listed beam parameters can cause the VDF to be unstable, the fraction of intervals where the beams are the major emitting component is significantly increased, as shown in \emph{panels (d1)} and \emph{(d2)}. 
As the beam detection instrumental effects become the dominant driver for the fraction of unstable measurements with a detected proton beam at $N_{C(cc)} > 0.4$, all observations from this region will be excluded from further consideration. 

The second issue, related to potentially undetected beams playing a role in the instability analysis at $N_{C(cc)} \sim 0.1$ is more complicated.  
Namely, even though $\Delta v_{b,c} / v_{A,c}$ decreases with both  $N_{C(cc)}$ and $r$ \citep{Alterman_2018_ApJ}, beam parameters could still have non-negligible values in this region, and non-detection of beams could significantly after the predicted plasma response. 
The manifestation of undetected beams influence is illustrated in \emph{panels (d2)} and  \emph{(d4)} of Figure \ref{fig:overview_N_C}. 
If the proton beam is detected, then we observe a constant trend transferring from the core being the dominant free energy source in the young wind towards it's irrelevance as a source of free energy as it isotropizes \citep{Kasper_2017}. 
On the other hand, some of the cases where only core and $\alpha$ components are identified will have an unresolved beam embedded into the core. 

The most important effect of this missed detection is the change, either increase or decrease, of the fitted parallel temperature component of the core. 
Here, we define the effective beam parallel temperature in terms of the particle kinetic energy in the reference frame of the core
\begin{equation}
  T_{\mathrm{eff}\parallel,b} = T_{\parallel,b} + \frac{m_p (\Delta v_{b,c})^2}{2 k_b}
  \label{eq:T_b_eff}
\end{equation}
and the joint effective parallel temperature of core and beam as
\begin{equation}
  T_{\mathrm{eff}\parallel} = \frac{n_c T_{\parallel,c} + n_b T_{\mathrm{eff}\parallel,b} }{n_c + n_b},
  \label{eq:T_eff}
\end{equation}
being the parallel temperature of the proton VDF when the beam is not detected. 

\begin{figure*}
  \includegraphics[width=\textwidth]{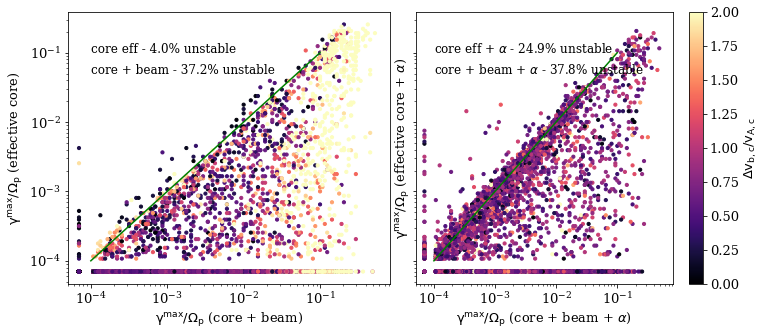}
  \caption{Fastest growth rates with effective core parallel temperatures compared to observed temeprature with resolved core and beam, for cases where $\alpha$ population was detected (\emph{right}), or was either undetected or removed from the analysis (\emph{left}). Stable intervals are shown with an imposed value of $\gamma^{\textrm{max}}/\Omega_p = 7 \cdot 10^{-5}$. }
  \label{fig:effective_gamma}
\end{figure*}

The impact of $T_{\parallel,c}$ approaching $T_{\mathrm{eff}\parallel}$ depends on where in the phase space it occurs. 
In the young wind, core anisotropy is high and an artificial increase in the parallel component pushes the $T_{\perp,c} / T_{\parallel,c}$ ratio towards unity, making the modelled VDF less prone to instabilities triggered by anisotropy, such as ion-cyclotron (IC) and mirror instability. 
Also, the core stops being a completely dominant emitting component in this region, but there is a notable fraction of unstable intervals that are driven by very hot $\alpha$ particles \citep{Stansby_2019_A&A}. 
On the contrary, older wind features $T_{\perp,c} / T_{\parallel,c} \sim 1$, and sufficient increase in the parallel temperature can make the core susceptible to firehose (FH) instability due to an increase in the total parallel pressure. 
This causes the core to maintain its contribution to the unstable behavior of plasma when beams are undetected, which is diminished when the beams are detected. 
The impact of the effective decrease in $T_{\mathrm{eff}\parallel}$ are opposite of the ones described. 

Since failure to detect beams is more likely to happen for lower densities and drifts, we address the case of intermediate instead of low Coulomb numbers in more detail. 
It is worth noting that, since we have no knowledge of a potential beam when it is not resolved, we will use the intervals that have a beam, and then evaluate what the effects on the linear stability would be if that beam was not detected, but rather treated as a part of the core VDF. 
We focus on the observations with $0.3< N_{C(cc)} <0.4$ (black shaded area on \emph{panel (d4)}). 
In this region, there is a total of $77,148$ fitted VDFs, and a subset of $42,033$ that contain beam components. 
Within the intervals of interest, $14,628$ VDFs also contain $\alpha$ components. 
First, we calculate $T_{\mathrm{eff}\parallel}$ for each VDF and reevaluate $\mathcal{P}_{0(eff)}$ and $\mathcal{P}_{eff}$ based on that value.
Then, we reanalyze these intervals using the procedure described in Section \ref{ssec:analysis}. We perform the instability analysis for intervals with $\alpha$ components two times, both including $\mathcal{P}_{\alpha}$ and with the helium artificially removed in order to isolate the undetected beam effects. 

\begin{figure*}
\centering
  \includegraphics[width=0.6\textwidth]{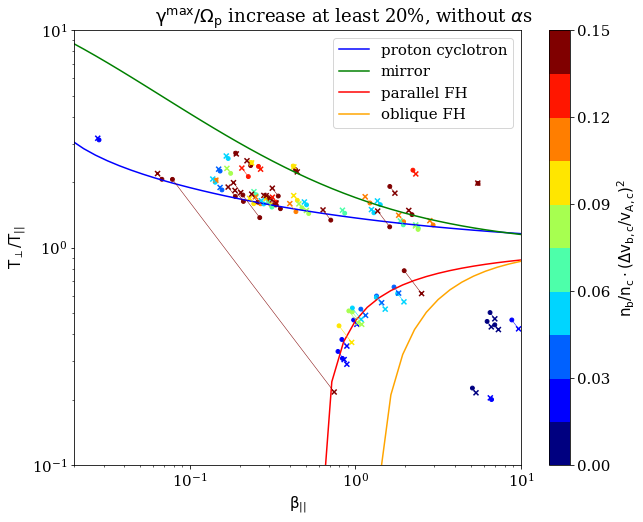}
  \caption{The shift of the core population in the $\big(\beta_{||,c}, T_{\perp,c} / T_{||,c} \to \beta_{||,\mathrm{eff}}, T_{\perp,c} / T_{\mathrm{eff}\parallel} \big)$ space due to possible non-detection of beams, with dots and X marks representing C+B and effective case, respectively. 
  The instability thresholds are calculated by \cite{Verscharen_2016_ApJ}. 
  Only intervals where the maximum growth rate is increased in the effective case by at least 20\% are shown. }
  \label{fig:brazil_shift}
\end{figure*}

The results are shown on Figure \ref{fig:effective_gamma}. For each interval, the two scenarios are compared---having the core and beam as observed (further on, referred to as "C+B case"), and having core and beam replaced with a single proton component with $T_{\mathrm{eff}\parallel}$ ("effective case"). The contribution of the beam to the perpendicular pressure is not enhanced by the drift, and we approximate $T_{\mathrm{eff}\perp} \approx T_{\perp,c}$. As $n_b \ll n_c$, a correction to this value that would arise from a more sophisticated approach that takes into account the difference between $T_{\perp,c}$ and $T_{\perp,b}$ would be very small. 
Interactions of unstable modes with $\alpha$ particles can be very complicated (see Section \ref{ssec:Radial_Trends} and also \cite{Klein_2019_ApJ}), but even the cases that include identified helium (\emph{right panel}) show an intuitive general trend. 
The effective case is generally more stable than the C+B case, as the sensitive beam structure and related velocity space gradients are is removed. 
This effect is clearly seen at the \emph{left panel}, where C+B cases with strongly drifted beams have $\gamma^{\textrm{max}}/\Omega_p$ increased by one to three orders of magnitude compared to their effective counterparts. 
However, some of the intervals, usually the ones with low beam drifts, tend to be more unstable in the effective case. 
This unusual behavior can have two causes, as either 1) the beam drift is very low, so it does not trigger any unstable modes, but it contains enough phase space density to potentially "push" the core population over the marginal stability threshold in the effective case; or 2) the beam is significantly drifted but has very low density, and is not as strongly unstable as if it's contribution to the parallel pressure would be added to the core.

For the beam detection issues, only first scenario is of interest, as only beams of $\Delta v_{b,c} / v_{Ac} \leq 0.5$ have realistic possibility to remain undetected. 
Only $119$ (0.28\%) effective VDFs from this subset tend to either be unstable when its equivalent C+B distribution is not, or have higher $\gamma^{\textrm{max}}/\Omega_p$ if both are unstable, so there will not be a significant number of cases where unphysical highly unstable modes are inferred. On the opposite end, we find that possible non-detection of the beam could "pacify" unstable VDFs, as $13,942$ (33.15\%) of VDFs are unstable in C+B case, but are stable in the effective case. 
However, only $1,177$ (2.8\%) are cases with $\Delta v_{b,c} / v_{Ac} \leq 0.5$. Therefore, we proceed to keep the intermediate $N_{C(cc)}$ intervals in our analysis. 

To further illustrate this point, we show the difference between C+B (core) and effective VDFs on $\beta$-anisotropy diagram shown on Figure \ref{fig:brazil_shift}. 
Only cases where $\gamma^{\textrm{max}}/\Omega_p$ is increased by at least 20\% between C+B and effective case are shown for clarity, and are colored with respect to $n_b / n_c \cdot \Delta v_{b,c}^2 / v_{Ac}^2$, which is used as a proxy for a contribution to the parallel pressure. 
This quantity is almost completely determined by $\Delta v_{b,c}^2 / v_{Ac}^2$ (not shown). 
Here, we again emphasize that a hypothetical undetected beam can work two ways---total parallel pressure can either increase or decrease, depending on the value of $T_{\mathrm{eff}\parallel,b}$ (Equation \ref{eq:T_eff}), which can be lower than $T_{\parallel,c}$ if the drift is low. 
For $T_{\mathrm{eff}\parallel,b} < T_{\parallel,c}$, the core population is shifted to the left on the plot, with effective increase in anisotropy, and if the VDF is close to the IC instability threshold, the value of $\gamma^{\textrm{max}}/\Omega_p$ can be overestimated. 
A similar scenario can happen if the VDF is close to FH threshold and the effective parallel pressure is increased due to $T_{\mathrm{eff}\parallel,b} > T_{\parallel,c}$. From Figure \ref{fig:brazil_shift} is visible that, in general, the shift in the parameter space is not large if the drift is not large. 
As strongly drifted beams have a significantly lower chance to be undetected in this part of phase space, an eventual mistreatment of slightly drifted beams will cause only small changes in the estimated maximum growth rates, while intervals that are incorrectly estimated to be unstable due to this mistreatment are only those that already are very close to marginal stability thresholds. 
This adds another level of confidence to our approach of treating the intermediate $N_{C(cc)}$ observations as sufficiently reliable.

\subsection{Dependencies on Solar Wind Parameters}
\label{ssec:Radial_Trends}

\begin{figure*}
  \includegraphics[width=\textwidth]{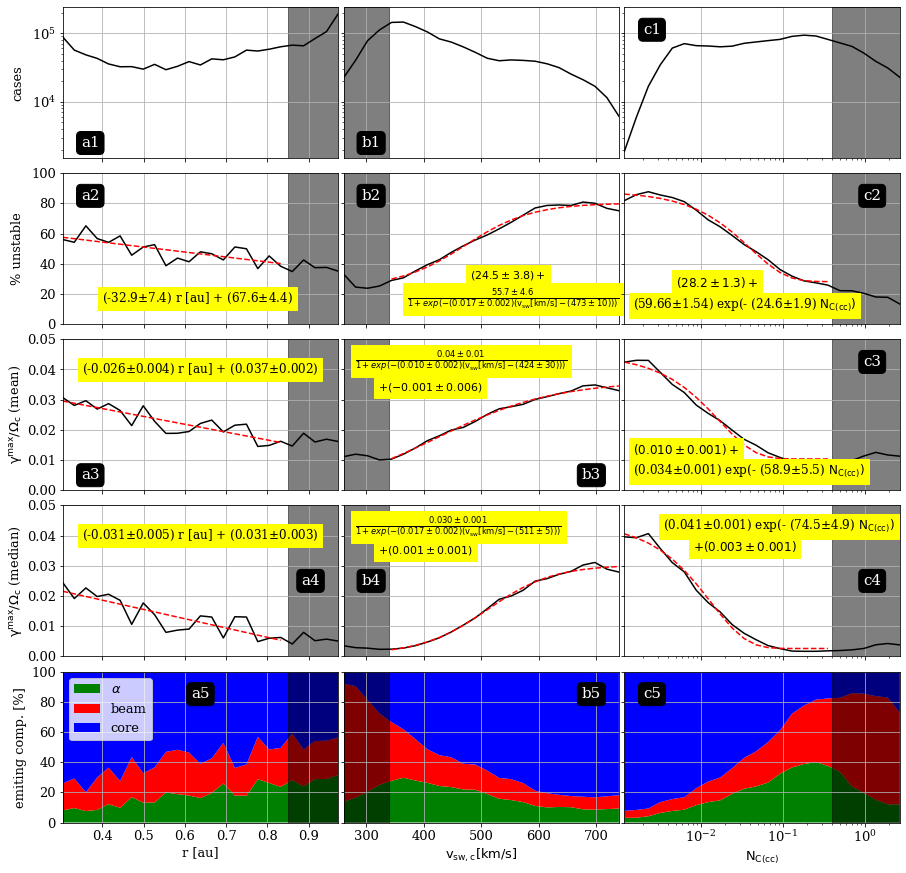}
  \caption{Fitted linear and exponential dependencies of the fraction of unstable intervals \emph{(b)}, maximum growth rate mean \emph{(c)} and median \emph{(d)}, from radial distance \emph{(1)}, solar wind speed \emph{(2)} and Coulomb number \emph{(3)}. 
  \emph{Panels (1)} and \emph{(5)} are showing the number of sampled intervals and the fractions of the emitting VDF components, respectively. 
  Black shaded regions are removed from consideration based on instrumental limitations explained in Section \ref{ssec:Data_Caveats}. }
  \label{fig:1d_hist}
\end{figure*}

We summarize the dependencies of the unstable behavior on $r$, $v_{sw}$ and $N_{C(cc)}$ in Figure \ref{fig:1d_hist}. 
The intervals excluded based on the analysis described in Section \ref{ssec:Data_Caveats} are shaded in black. 
We restrict our data analysis as a function of radial distance and solar wind speed by identifying the unexpected rise in the mean and median values of the maximum growth rate and, for the case of $v_{sw,c}$, also a sudden increase in participation of proton beam as the emitting component. 

The linear trends on the \emph{panels (a2-4)} are showing that, despite the variation of plasma parameters measured at each radial distance, we still have a steady decline in both instability occurrence and growth rates. 
The fraction of unstable intervals is steadily growing as we approach the Sun, with linear extrapolation aiming at approximately two thirds of all intervals being unstable in the corona. 
It is unclear from this data set how close to the Sun this trend is maintained. 
The differences between line slopes for mean and median values of $\gamma^{\textrm{max}}/\Omega_p$ are of the order of 20\%, which is not a large value for a quantity with spread of three and a half orders of magnitude. 

The inferred dependencies become significantly more extreme if the results are organized with respect to $v_{sw,c}$ or $N_{C(cc)}$. 
From profiles of the emitting components (\emph{panels (5)}) is clear that these two ways of organizing the data have many similarities, which are intuitively expected as $N_{C(cc)}$ roughly scales as $v_{sw}^{-1}$ (\emph{panel (d1)} of Figure \ref{fig:overview_N_C} is shown again for the sake of completeness). 
For both parameters, the instability growth and occurrence rates show an exponential rise or decay between maximum and minimum values, both of which seem to be separated by a marginally larger amount for the case of $N_{C(cc)}$. 
For both underlying variables, the difference between mean and median values of $\gamma^{\textrm{max}}/\Omega_p$ demonstrates similar behavior, with less than 20\% discrepancy close to the maximum. 
Then, during the phase of exponential decay, the mean-to-median difference jumps to a factor of $\sim 5$, which is quantified by the zeroth order terms in the fitted exponential functions. 
The described behavior is matched with the decay of the core population as the main source of free energy, and beams and helium gradually taking over that role. 
At low speeds the core is not as dominant as at low Coulomb numbers, and is also not as insignificant at high speeds as at high $N_{C(cc)}$. 
It also follows the pattern shown on Figure \ref{fig:overview_N_C}, \emph{panel (e)}, where the power emitted by the core is mostly absorbed by $\alpha$ particles at the collisionally young wind. 
As the core becomes isotropic, beams are more likely to be the main driver of instability, and occasionally some of the emitted power is absorbed by the core. 
A similar graph is produced if the data is organized by $v_{sw,c}$ (not shown), but the trends are slightly less pronounced, similar as on the \emph{panels (b5)} and \emph{(c5)} of Figure \ref{fig:1d_hist}. 

Aside from the Coulomb number, which depends only on in situ characteristics of the plasma, a parameter frequently used in the literature is the collisional age $A_c$, the integral of all the Coulomb collisions experienced by a parcel of plasma as it propagates outwards from the Sun's surface (see e.g. \cite{Maruca_2013_PhRvL,Kasper_2017}). 
We reprocessed the results on Figures \ref{fig:overview_N_C} and \ref{fig:1d_hist} in terms of $A_c$, as well as different variations of Coulomb number that estimate collisions of core population with beams $N_{C(cb)}$ and helium $N_{C(c\alpha)}$. 
Results obtained are not visually distinguishable form the ones shown here, but subtle differences between these various collision "clocks" still could exist, and their effect on properties of unstable modes will be examined in detail in the follow-up article. 


\section{Discussion and Conclusions}
\label{sec:discussion}

The database from 15 years of \emph{Helios} observations of ion VDFs processed by the \texttt{PLUME} dispersion solver provides sufficiently robust statistics of the inferred behavior of linear instabilities between 0.3 and 1 au. 

Linear trends shown on \emph{left panels} of Figure \ref{fig:1d_hist} are different from the results shown by \cite{Klein_2019_ApJ}, who found a constant fraction of unstable intervals to be over 80\% at all radial distances. 
As that work used only single core and helium populations overwhelmingly sampled in the fast solar wind with $v_{sw,p}>600 \mathrm{km/s}$, the reasons for this discrepancy can be seen in \emph{middle panels}. 
For very high speeds, both fraction of unstable modes and $\gamma^{\textrm{max}}/\Omega_p$ means and medians are reaching maximum values marginally lower than the ones presented by \cite{Klein_2019_ApJ}, probably due to the selection bias of that work in using only clearly distinguished components, which naturally prefers intervals with more extreme parameters and also could occasionally misidentify beam contribution (Section \ref{ssec:Data_Caveats}). 

The exponential decay in mode activity signals that very slow, and also highly collisionally processed solar wind is still unstable for about a quater of intervals, but the intensity of the wave fluctuations is significantly decreased. 
This could be the indicator of core population moving away from the IC and mirror instability thresholds and the plasma becoming more isotropic at larger radial distances \citep{Matteini_2007_GRL}. 
Once the core is nearly isotropic and beam and helium drifts are sufficiently small, and also the value of $\beta_{\parallel}$ increases for all populations due to expansion, the plasma moves towards the FH instability threshold. 
The marginal stability could then be violated by a variety processes such as VDF elongation due to magnetic moment conservation \citep{Livi_1987_JGR}, long-wavelength compressive fluctuations \citep{Verscharen_2016_ApJ}, or plasma heating due to large amplitude fluctuations \citep{Chandran_2010} and wave resonances \citep{Kasper_2013,Chen_2019}. 

As a separate topic, it is worth discussing the power transfer between VDF components, shown on Figure \ref{fig:overview_N_C}, \emph{panel (e)}. 
Intervals where the stable components absorbs significant fraction of the emitted free energy are not frequent, not surpassing 6\% at any $N_{C(cc)}$, but are a useful indicator of conditions at which each particular unstable mode resonates with any of the components. 
As mentioned above, the core is very anisotropic in the collisionally young solar wind and acts as the main source of free energy. The main unstable modes triggered are the Alfv\'en (IC) and entropy (mirror) mode. 
Neither of these modes has high resonant velocity \citep{Gary_1993}, and does not interact with strongly drifted beam particles \citep{Daughton_1998_JGR,Daughton_1999_JGR}. 
Helium has notably smaller drifts in the young wind \citep{Kasper_2017,Durovcova_2019_SoPh}, and therefore has more particles that can resonate with the growing modes (light blue shade). 
When the core anisotropy drops, the less populated components take over the role of the primary free energy source. 
Also, beam drifts decrease gradually, getting close enough for the excited IC and fast magnetosonic modes to interact first with $\alpha$ (pink) and then with the core particles (tomato red). 

Finally, although instrument limitations (Section \ref{ssec:Data_Caveats}) limit the confidence in our data close to 1 au, we still can conclude that all the components are moving towards stability in approximately linear rate with radial distance. 
As there are no significant energizing mechanisms that increase in effectiveness around 1 au, we hypothesize that the linear instabilities are least abundant in the solar wind at 3-5 au, as collisional processing moves the VDF away from the FH instability threshold \citep{Bale_2009_PRL,Matteini_2013_JGRA}. Further away, energy sources that originate outside of the Solar system, such as pick up ions, could start playing a role in the energy balance \citep{Goldstein_1995_GRL}. 

While we elaborated on the general trends of instability growth rates and consequences of technical limitations of I1a instrument, a detailed examination of the physical aspects is left for future work, in which we will investigate the dynamics of interaction between collisions and instabilities, and offer the general model of the instability behavior in the inner heliosphere. 


\begin{acknowledgments}
M. M. Martinovi\'c and K. G. Klein were financially supported by NASA grants 80NSSC19K1390 and 80NSSC19K0829.
K.G.K. is supported by NASA ECIP Grant 80NSSC19K0912. 
T. {\v{D}}urovcov{\'a} was supported by the Grant Agency of the Charles University under the project number 264220 and by the Czech Science Foundation under Contract 19-18993S. 
B. L. Alterman acknowledges NASA contract NNG10EK25C. 
An allocation of computer time from the UA Research Computing High Performance Computing at the University of Arizona is gratefully acknowledged.
\end{acknowledgments}



\end{document}